\documentclass[reprint,amsmath,amssymb,aps,prl,superscriptaddress]{revtex4-1}  
\usepackage{graphicx}  
\usepackage{dcolumn}   
\usepackage{bm}

\usepackage{color}   

\begin{document}

\preprint{APS/123-QED}

\title{
Spin decontamination for  magnetic dipolar coupling calculations: \\ Application to high-spin molecules and solid-state spin qubits} 

\author{Timur Biktagirov}
\author{Wolf Gero Schmidt}
\author{Uwe Gerstmann}
\affiliation{Lehrstuhl f{\"u}r Theoretische Materialphysik, Universit{\"a}t Paderborn, 33098 Paderborn, Germany}

\date{\today}%

\begin{abstract}
An accurate description of the two-electron density, crucial for magnetic coupling in spin systems, provides in general a
major challenge for density functional theory calculations.  
It affects, e.g., the calculated zero-field splitting (ZFS) energies of spin qubits in semiconductors that 
frequently deviate significantly from experiment. In the present work 
(i) we propose an efficient and robust strategy to correct for spin contamination in both extended periodic and finite-size systems,
(ii) verify its accuracy using model high-spin molecules, 
and finally (iii) apply the methodology to calculate accurate ZFS of spin qubits (NV$^-$ centers, divacancies) in diamond and silicon carbide. The approach is shown to reduce the dependence on the used exchange-correlation functional to a minimum.

\end{abstract}

\maketitle


An electron-electron magnetic dipolar interaction in a high-spin molecular system leads to the energy splitting of its 
$m_S$ spin sublevels in the absence of external magnetic fields \cite{Abragam, Harriman}. 
A theoretical prediction of the magnetic dipolar coupling and the resulting zero-field splitting (ZFS) is, however, challenged by 
the so-called spin contamination of the two-particle spin density \cite{Sinnecer_Neese, Jost}.
It arises in electronic structure calculations if the spin channels are allowed to differ in spatial representation (see also Fig.~1), 
leading to wave functions that are not eigenfunctions of the total spin-squared operator $S^2$  \cite{Baker, Ferre}.  
In particular, it can result in non-vanishing ZFS even for infinitely large spin separation \cite{Jost}.

Although the effects of spin contamination are well-known for molecular systems, 
they have scarcely been considered for solids. This is related to the fact that the remedy typically used to correct for 
spin contamination in density functional theory (DFT) calculations with a localized basis set, 
i.e., the the unrestricted natural orbital (UNO) approach \cite{Sinnecer_Neese},
cannot be straightforwardly realized in a plane-wave description of systems with periodic boundary conditions.
High-spin states in solids, however, attract significant attention as potential spin qubits, i.e. building blocks 
of quantum technological devices. Here, the ZFS serves as an essential lever to 
spectroscopically address and distinguish a specific spin center among other defects present in the host material.

\begin{figure*}[t]
\begin{center}
    \includegraphics[width=1.0\textwidth]{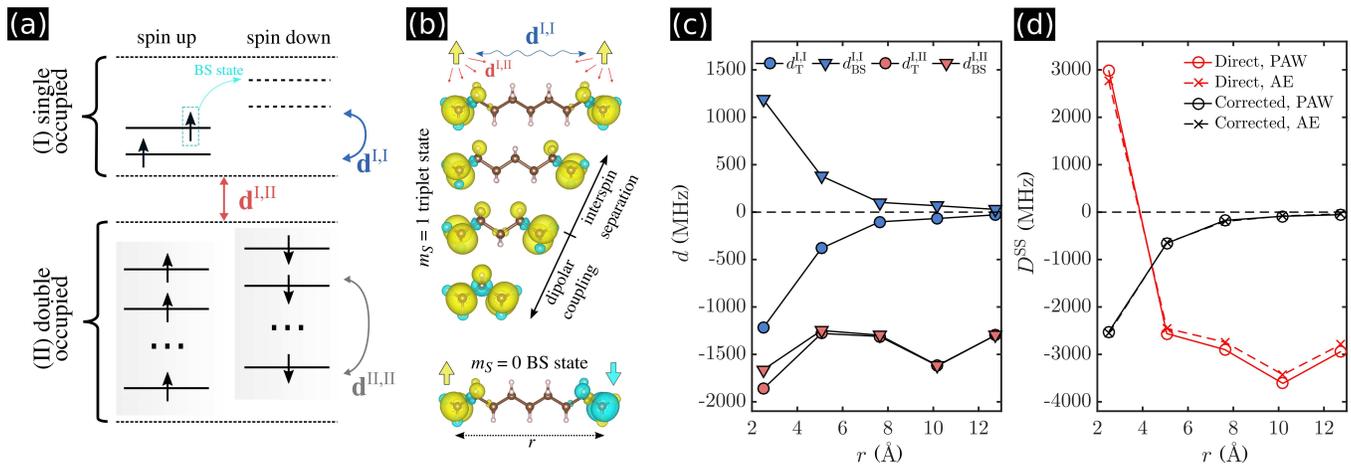}
\vspace*{-0.6cm}
\caption{\label{fig1} (a) Schematic illustration of the electron-electron magnetic dipole interaction in a spin-triplet system.
The overall coupling $\bold{d}$ originates from the interaction between the two single occupied
Kohn-Sham orbitals (denoted as I), as well the contributions arising from the double occupied orbitals (II). 
(b) A set of biradicals of different length (general formula C$_n$H$_{2n}$; adopted from Ref. [\onlinecite{Jost}]). 
Magnetization density in their triplet and BS states is illustrated with yellow (positive) and cyan (negative) isosurface.
(c) The contributions to the magnetic dipolar coupling in the triplet ($d_{\rm T}^{\rm I,I}$, $d_{\rm T}^{\rm I,II}$) and BS ($d_{\rm BS}^{\rm I,I}$, $d_{\rm BS}^{\rm I,II}$) states of the biradicals.
(d) The $D^{\rm SS}$ values obtained without the spin-contamination correction ({\it red circles}) and with the corrected approach ({\it black circles}). 
For comparison, the results of all-electron (AE) DFT calcuations (\texttt{ORCA} software \cite{ORCA}, PBE functional \cite{PBE} 
and def2-TZVP basis set \cite{TZVP}) with the direct ({\it red crosses}) and UNO-corrected ({\it black crosses}) approaches are presented.
}
\end{center}
\end{figure*}

In the present work we propose an efficient strategy for spin decontamination in magnetic dipolar coupling driven ZFS, which is
suitable for extended periodic systems. It is based on taking both the  $m_S = S$ and $m_S = S-1$ spin states of the system into account when 
calculating the spin-dipolar interaction.
The approach is first demonstrated using a molecular biradical as model system and subsequently applied to rectify the calculated ZFS
 of the nitrogen-vacancy (NV$^-$) \cite{Csore_NV_2017, Bardeleben_NVV_2016, Gerstmann2016, Sato_NV_2019, Zargaleh_NV_2018,Wang_2019},
 silicon-carbon divacancy (VV$^0$) \cite{Koehl_VV_2011, Falk_VV_2013, Casas_VV_2017, Christle_VV_2017, Seo_VV_2016, Zwier_VV_2015} and silicon vacancy (V$_{\rm Si}^-$) \cite{Soltamov_PRX_2016, Ivady_2017,Wrachtrup_NatureMat_2015,Soltamov_NatPhys_2014}
 centers in silicon carbide (SiC), and the NV$^-$ center in diamond \cite{Doherty_review, Awschalom_2018, Yao, Hensen, Arai, Dolde, Kucsko, Doherty_pressure, Barson, Ivady_2014}. 
Finally, we demonstrate that the established strategy can generally be applied for $S\ge1$ systems and, moreover, is robust with respect to the exchange-correlation functional.

The phenomenological spin Hamiltonian \cite{Abragam}
\begin{equation}
\hat{\mathcal{H}}^\textrm{ZFS} =  \hat{\bold{S}} \cdot \bold{D} \cdot \hat{\bold{S}}
\end{equation}
is conventionally adopted to characterize the ZFS in terms of the parameters 
$D = D_{zz}-\frac{1}{2}(D_{xx} + D_{yy})$ and  $E = \frac{1}{2}(D_{yy} - D_{xx})$, where $D_{ii}$ are the eigenvalues of the symmetric and traceless $3\times 3$ tensor $\bold{D}$, and
axially symmetric $\bold{D}$ tensors lead to vanishing $E$.

Within the DFT framework, the spin-dipolar coupling driven $\bold{D}$ tensor (referred to as spin-spin ZFS) 
is derived from the spatial distribution of Kohn-Sham 
orbitals obtained with standard spin-polarized self-consistent field 
calculations \cite{McWeeny1961, vanWullen_ZFS, Rayson_Briddon, Biktagirov, Bodrog_Gali}.
For the $S\ge1$ high-spin state of a system, the spin-spin $\mathbf{D}$ tensor is given by 
\begin{equation}
\label{eq1}
\begin{array}{c}
 \displaystyle
\bold{D}^{\rm SS}  = \frac{\bold{d}}{S(S-\frac{1}{2})}   
\:,
\end{array}
\end{equation}
where the elements of the traceless $3\times3$ matrix $\bold{d}$ are constituted by magnetic dipolar interaction between all 
the pairs of occupied Kohn-Sham states $m$ and $n$ (with $a,b = x,y,z$): 
\begin{equation}
\label{eq2}
\begin{array}{c}
 \displaystyle
d_{ab} = \frac{\alpha^2}{8}\sum_{\substack{m, n} }\chi_{m,n} 
\int 
\frac{|\mathbf{r} - \mathbf{r'}|^2\delta_{ab}-3(\mathbf{r} - \mathbf{r'})_{a}(\mathbf{r} - \mathbf{r'})_{b}}{|\mathbf{r} - \mathbf{r'}|^5} 
\\ 
\\
 \displaystyle
\ \times \big[n_{mm}(\mathbf{r}) {n_{nn}^*(\mathbf{r'}) - n_{mn}(\mathbf{r})} {n_{mn}^*(\mathbf{r'})}  \big] ~d\mathbf{r} d \mathbf{r'} 
\:.
\end{array}
\end{equation}
In Eq.~\ref{eq2}, $\alpha$ is the fine-structure constant, the charge densities $n_{mn}(\mathbf{r})=\psi_m(\mathbf{r})\psi_n(\mathbf{r}) $ 
reflect the spatial distribution 
of the orbitals, and $\chi_{m,n}$ originates from the matrix elements of spin-operators: $\chi_{m,n}=1$ when the 
orbital belong to the same spin channel, and $\chi_{m,n}=-1$ otherwise. Throughout the manuscript, we will also make use of the magnetic dipolar coupling parameter 
$d = d_{zz}-\frac{1}{2}(d_{xx} + d_{yy})$ analogous to the $D$ value. Note that the expression in the square brackets in 
Eq.~\ref{eq2} is an approximation of the two-particle spin density which is, otherwise, not directly available from DFT.

Due to spin polarization, the Kohn-Sham states of a paramagnetic system have different energies and different spatial 
distributions in the spin-up and spin-down channels. Therefore, the spin-spin ZFS is not entirely determined by the 
coupling between the half-filled states (denoted as $\bold{d}^{\rm I,I}$ in the energy level diagram in Fig.~1a, where a
 spin-triplet $S=1$ case is shown as a prototype example).
Instead, it also contains non-vanishing collective contribution from the two-particle spin densities of single-/double- 
($\bold{d}^{\rm I,II}$) and double-/double-occupied ($\bold{d}^{\rm II,II}$) states.
Note that in the absence of spin polarization, the contributions from the spin-up and spin-down electrons of the same 
double-occupied orbital would cancel out due to the spin-dependent factor $\chi_{m,n}$. 

The spin contamination manifests itself in unphysical spin-dipolar interaction which induces spurious 'on-site'
terms in the two-particle spin density \cite{Jost} entering $\bold{d}^{\rm I,II}$.
This can be nicely demonstrated by adopting the family of $S=1$ biradicals used in Ref.~\onlinecite{Jost} as a
reference system (see Fig.~1b). Here, the unpaired electrons are located at the opposite termini, so that the spin-spin ZFS
{\em must} decrease according to the point-dipole approximation as the molecule gets longer and longer. This is, however, not the case as 
the unphysical 'on-site' contributions 
($\bold{d}_{\rm T}^{\rm I,II}$ in Fig.~1c, where the subscript is introduced to distinguish the triplet spin state)
result in unrealistic $D^{\rm SS}$ values (Fig.~1d).

The considered biradicals represent a lucky case where the physical and spin contamination driven parts of the magnetic
dipolar coupling tensor are well-separated. Indeed, for the longer molecules, the coupling is completely described by
$\bold{d}_{\rm T}^{\rm I,I}$, while $\bold{d}_{\rm T}^{\rm I,II}$ is being entirely caused by spin contamination.
In general, there is, however, no evidence that the spin contamination can be attributed only to selected terms in
Eq.~\ref{eq2} or that it can be treated as an additive error. Therefore, it must be excluded in a more elaborate way.
In this paper, we establish a systematic correction scheme based on the simple
idea that the spin-contamination error in $\bold{d}_{(m_S = S)}$ of the $m_S = S$ high-spin state 
of a general $S\ge1$ system is one-to-one reflected by its $m_S = S-1$ state. 
In Fig.~1c, this is illustrated for the prototypical $S=1$ case. Its $m_S = 0$ low-spin configuration 
is a single-Slater-determinant state and thus
should be referred to as the broken-symmetry, BS, state.
As a consequence, the average $\overline{\bold{d}}_{(m_S = S-1)}$ among all the possible $m_S = S-1$ configurations 
restores the physical part while cancelling out the spin-contamination error completely (see also the discussion in the Supplemental Material (SM) \cite{SI}):
\begin{equation}
\label{eq3}
 \begin{array}{c}
\displaystyle
\tilde{\bold{D}}^{\rm SS} = \frac{\bold{d}_{(m_S = S)}-\overline{\bold{d}}_{(m_S = S - 1)}}{2S-1}
\:,
\end{array}
\end{equation}
Each of these $m_S = S-1$ configurations can be obtained by changing the occupation of one of the half-filled Kohn-Sham orbitals from spin-up to spin-down and subsequently performing the self-consistent field calculation.
Note that the $S$ dependent denominator in Eq.~4 is exactly the prefactor which relates the $D$ values and the energetic distance between the $m_S = S$ and $m_S = S-1$ spin sublevels \cite{SI}.

For an $S=1$ system, this follows directly from Fig.~1c. While showing no net spin, its $m_S = 0$ BS state 
exhibits antiferromagnetic coupling between the spin-up and spin-down electrons occupying two half-filled orbitals, which leads
to non-vanishing $\bold{d}_{\rm BS}^{\rm I,I}$ ({\it blue triangles} in Fig.~1c). 
Thus, in a '{\em perfect}' spin-triplet system, i.e. a system {\em without} spin contamination,
$ \bold{d}_{\rm BS} = - \bold{d}_{\rm T}$ is expected. 
Consequently, when the spin-spin ZFS of a  triplet system is defined as 
$\tilde{\bold{D}}^{\rm SS} = (\bold{d}_{\rm T}-\bold{d}_{\rm BS})/2$,
the {\it physical} $\bold{d}_{\rm T}^{\rm I,I}$ part is conserved, while the spin contamination error of the triplet and BS states cancel.

As an illustration, we apply the proposed strategy to the considered set of biradicals. 
Here and in the following, the spin-spin ZFS calculations \cite{Biktagirov} are based on the projector augmented wave (PAW) formalism \cite{Blochl_paw} in combination with norm-conserving pseudopotentials \cite{Troullier_Martins}, 
as implemented in a modified version of the GIPAW module of the \texttt{Quantum ESPRESSO} software \cite{Giannozzi_2009, Giannozzi}. 
We used the PBE exchange-correlation functional~\cite{PBE},
and a plane-wave (PW) basis set with 700 eV kinetic energy cutoff. We find that the corrected ZFS 
exhibits perfect agreement with the UNO-based all-electron DFT results (Fig.~1d). Most importantly, in contrast to the
UNO approach, the established scheme is applicable for high-spin systems with periodic boundary conditions.

\begin{figure}[t]
\begin{center}
    \includegraphics[width=1\columnwidth]{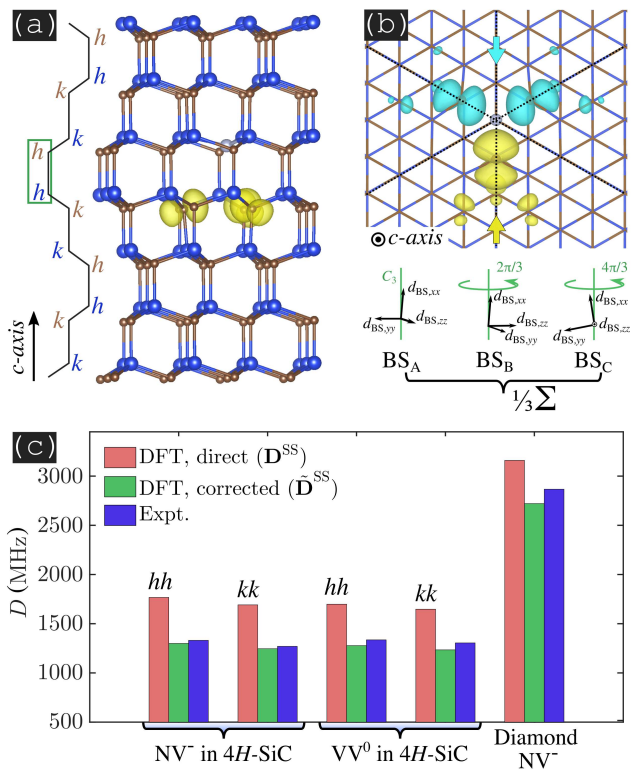}
\caption{\label{fig2} Magnetic dipolar coupling driven ZFS of the NV$^-$ and VV$^0$ divacancy centers. 
(a) Crystal structure of $4H$-SiC illustrating the two inequivalent Si ({\it blue}) and C ({\it brown}) lattice sites.  
Spin density of the NV$^-$ center with the axial $hh$ configuration is shown ({\it yellow isosurface}).
(b) Spin density distribution in a selected BS state of the axial NV$^-$ center and ({\it below}) 
schematic illustration of the three possible $\bold{d}_{\rm BS}$ tensor directions. 
The 'true low-spin' $\overline{\bold{d}}_{(m_S = 0)}$ tensor is defined as their average restoring the symmetry of the triplet state (in case 
of axial pairs, $C_{3v}$). 
(c) The spin-spin $D$ values obtained without ({\it red}) and with ({\it green}) the proposed spin contamination correction in comparison 
with the experimental data from Refs. [\onlinecite{Csore_NV_2017, Gerstmann2016, Falk_VV_2013, Ivady_2014}] ({\it blue}).
}
\end{center}
\end{figure}

Subsequently, we apply the correction to the NV$^-$ and VV$^0$ defects in $4H$-SiC, and the NV$^-$ center in diamond. 
These spin-triplet defect centers are arguably among the most prominent solid-state spin qubits. 
Their ZFS is believed to be almost entirely caused by magnetic dipolar coupling 
and has been successfully treated with plane-wave pseudopotential DFT \cite{Gerstmann2016, Ivady_2014, Csore_NV_2017}. 
Some recent publications \cite{Seo, Biktagirov, Ivady_npj}, however, argued that the reported agreement between 
the calculated and the measured ZFS profits from error cancellation due to incomplete treatment of the 
pseudized electron density. Indeed, as shown in Ref. [\onlinecite{Biktagirov}], the resulting ZFS depends 
significantly on the technical parameters adopted for the pseudopotentials generation, unless the '{\em true}' 
character of the wavefunction within the atomic core region is reconstructed.
Such reconstruction can be achieved by the PAW method \cite{Blochl_paw}. 
As pointed out by Iv{\'a}dy et al. \cite{Ivady_npj}, the fully PAW reconstructed ZFS of the spin-triplet centers in the SiC polytypes is, however, 
almost 30\% larger than the values reported experimentally.

In diamond, there is only one magnetically distinct NV$^-$ center. It can be aligned with one of 
the $\big<111\big>$ diagonals and always exhibits $C_{3v}$ symmetry.
In contrast, the hexagonal $4H$ polytype of SiC features a periodic sequence of quasicubic ($k$) and hexagonal ($h$) 
Si-C double layers along the crystal $c$-axis, offering inequivalent crystallographic positions and configurations 
for the defect pairs. 
Those NV$^-$ and VV$^0$ centers that are aligned with the hexagonal $c$-axis (the so-called {\em axial} pairs) 
have $C_{3v}$ symmetry (see Fig.~2a). 
For all other orientations, the symmetry is lowered towards $C_{1h}$.
Irrespective of the host material and the crystallographic position, the 
NV$^-$ and VV$^0$ defects share almost the same $m_S = 1$  electronic structure with two unpaired electrons either populating 
a double degenerate $e$ orbital (axial pairs) or slightly split $a'$ and $a''$ orbitals (basal pairs). 

As shown in Fig.~2c and Table I, even for the axially symmetric NV$^-$ and VV$^0$ centers, 
the ZFS depend on the lattice site sensitively. 
Notably, our results for SiC reproduce the previously reported 30\% discrepancy between the measured $D$ 
values and $D^{\rm SS}$, if directly calculated via Eq.~\ref{eq1}. 
It should be also mentioned that the coupling of the two half-filled Kohn-Sham states (i.e., the $\bold{d}_{\rm T}^{\rm I,I}$ term) 
explains in this case less than 70\% of the measured value, rendering a spin-restricted approach (where the same wavefunctions 
are used for spin-up and spin-down) to be insufficient.  The $\bold{d}_{\rm T}^{\rm I,II}$ and $\bold{d}_{\rm T}^{\rm II,II}$ 
terms are, thus, significant for NV$^-$ and must be retained in a systematic manner. 
In the following, we show that this can be accomplished by using the proposed methodology.

\begin{table}[t]
\caption{\label{tab:table1} The spin-spin ZFS for the NV$^-$ center in diamond, and NV$^-$, VV$^0$,  V$_{\rm Si}^-$ centers in 
$4H$-SiC calculated without (second column, $\bold{D}^{\textrm{SS}}$) and with (third column, $\tilde{\bold{D}}^{\rm SS}$)  spin contamination correction 
in comparison with experiment.\footnote{Experimental values taken from Ref.~\onlinecite{Ivady_2014} (diamond), 
Refs.~\onlinecite{Gerstmann2016, Csore_NV_2017} (NV$^-$ in SiC), Ref.~\onlinecite{Falk_VV_2013}  (VV$^0$), and from Ref.~\onlinecite{Ivady_2017} (V$_{\rm Si}^-$). 
The calculations performed using 512-atom (diamond) and 432-atom ($4H$-SiC) supercells with shifted 2$\times$2$\times$2 $k$-point grids.}
The ZFS are provided as the $D$ values ($E$ values in parentheses if non-zero).
}
\begin{ruledtabular}
\begin{tabular}{lccc}
Defect/site & $\bold{D}^{\textrm{SS}}$  & $\tilde{\bold{D}}^{\rm SS}$    & Expt. \\  
\hline
\multicolumn{4}{c}{axially symmetric $S=1$ defects } \\
\hline
Diamond NV$^-$	&3160.0 	&2720.7	&2867  \\   
NV$^-/hh$		&1767.3 	&1299.6	 &1331 \\  
NV$^-/kk$			&1691.6	&1245.7 	&1282 \\   
VV$^0/hh$		&1698.5 	&1277.6	 &1336 \\  
VV$^0/kk$		&1647.9	&1234.4 	&1305 \\   
\hline
\multicolumn{4}{c}{{\em basal}  $S=1$ defects} \\
\hline
NV$^-/hk$			&1614.8 	&1127.4	    &  1193  \\
				&(159.4)	&(120.9)     &  (104) \\  
NV$^-/kh$			&1733.6 	&1259.7	    &  1328  \\
				&(61.2)        & (8.9)     &  (15)  \\  
VV$^0/hk$		&1656.0 	&1219.2	    &  1334  \\  %
				&(42.2)	&(45.0)     &  (19) \\  %
VV$^0/kh$		&1592.3 	&1126.8	    &  1222  \\  %
				&(107.9)      & (89.3)     &  (82)  \\  %
\hline
\multicolumn{4}{c}{axially symmetric $S=3/2$ defects } \\
\hline
V$_{\rm Si}^-/h$	&26.5 	&1.8		 &2.6 \\  %
V$_{\rm Si}^-/k$	&47.4	&38.5 	&35.0 \\   %
\end{tabular}
\end{ruledtabular}

\end{table}

In contrast to the biradicals considered above (Fig. 1), the unpaired electrons of the {\em axial} NV$^-$ or VV$^0$ center 
are equally distributed among the three carbon dangling bonds surrounding the vacancy (cf. Fig, 2a).
Thus, in a $m_S =0$ configuration, the spin-up electron can localize at one of these dangling bonds, 
while the spin-down electron is being shared by the other two adjacent carbon atoms.
As illustrated in Fig.~2b, this provides three symmetry-reduced BS states of the axial defect. 
Thus, in order to apply Eq.~4 to the ZFS of NV$^-$ and VV$^0$, we define the  
$\overline{\bold{d}}_{(m_S = 0)}$ tensor as the $C_{3v}$ symmetrized average of the three resulting BS tensors. 

Figure 2c demonstrates that the spin decontamination applied in this way removes the overestimation and 
brings the calculated ZFS ($\tilde{D}^{\rm SS}$) in almost perfect agreement with the measured $D$ values. 
The improvement is especially decisive for the spin qubits in SiC.
Note that a part of the remaining discrepancies ($<$~100~MHz) can be attributed 
to second-order contributions to the ZFS due to spin-orbit coupling \cite{SO_ZFS}. 

The same procedure can be applied in case of the basal pairs in SiC. 
From Table~I it can be seen that these spin qubits suffer from the spin contamination to the same extent as their axially symmetric counterparts.
Averaging of their three no-longer equivalent BS configurations provides improved estimates for $D$, 
and also reasonable values for the rhombicity parameter $E$. 
Finally, we show that the proposed approach can be successfully used 
for higher spin qubits, i.e. the V$_{\rm Si}^-$ centers in 4$H$-SiC that possess a $S=3/2$ ground state.
Notably, the $D$ value of V$_{\rm Si}^-$ at the $h$ site is found to manifest enormous spin contamination error, which 
is eliminated by our correction scheme (cf. Table~I).
Because of an exceptional accuracy provided by our approach, the reported results
can be considered as a solid contribution that supports the assignment 
of the spectroscopic fingerprints observed in 4$H$-SiC as NV$^-$, VV$^0$ and V$_{\rm Si}^-$ centers.

\begin{figure}[t]
\begin{center}
    \includegraphics[width=0.8\columnwidth]{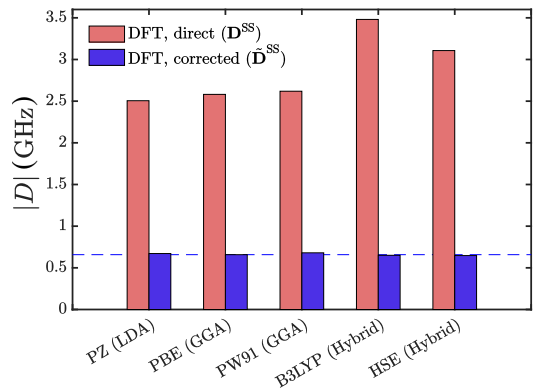}
\caption{\label{fig3} Spin-spin ZFS (the absolute value of $D$) of the C$_5$H$_10$ spin-triplet biradical (from the set considered in Fig.~1) 
calculated with different exchange-correlation functionals: PZ \cite{PZ}, PBE, PW91 \cite{PW91}, B3LYP \cite{B3LYP} and HSE \cite{HSE}. 
Horizontal dashed line indicating the $\tilde{D}^{\rm SS}$ value obtained with the PBE functional is provided as a guide for the eye.
}
\end{center}
\end{figure}

To summarize, in the present work we propose an efficient scheme to eliminate spin contamination in magnetic dipolar coupling. 
It allows us to identify the spin contamination of the two-electron density as a highly relevant source of discrepancy between measured and calculated ZFS of high-spin defects in semiconductors.  
This discrepancy can be lifted by calculating the difference of $m_S = S$ and $m_S = S-1$ magnetic dipolar coupling tensors.
This scheme is robust with respect to the exchange-correlation functional (see Fig.~3). 
In particular, it compensates the increased extent of spin contamination associated with the introduction of a fraction of exact Hartree-Fock exchange in hybrid functionals \cite{Baker, Menon, Neese_2007}. 
This opens up the possibility to use either hybrid functionals or (semi)local (LDA and GGA) functionals  while reaching the same level of accuracy for the spin-spin ZFS.
The proposed correction will be also beneficial for an accurate description of magnetic 
dipolar interaction between {\it distant} spin centers, 
e.g., for modelling a qubit coupled to a spin-bath \cite{Belthangady, Doherty_2016, van_Oort} 
or a solid-state spin probe coupled to surface electron spins \cite{Lukin_patent}.

\begin{acknowledgments}
Numerical calculations were performed using grants of computer time from the Paderborn Center for Parallel Computing (PC$^2$) and the HLRS Stuttgart. The Deutsche Forschungsgemeinschaft (DFG) is acknowledged for financial support via the priority program SPP 1601 and the Transregional Collaborative Research Center TRR 142 (project number 231447078). 
\end{acknowledgments}

\end{document}